\documentclass[letterpaper,english,twocolumn,prl, aps,showpacs,tightenlines]{revtex4-1}

\usepackage{dsfont}
\usepackage{amsfonts}
\usepackage{subfigure}
\usepackage{psfrag,graphicx}
\usepackage{dcolumn}
\usepackage{amsmath,amssymb}
\usepackage{amsbsy}
\usepackage{bm}
\usepackage{latexsym}
\usepackage{epstopdf}
\usepackage{epsf}
\usepackage{color}
\usepackage[english]{babel}
\usepackage{natbib}
\usepackage{appendix}

\usepackage{color}
\usepackage{soul}

\definecolor{mygrey}{gray}{0.35}
\definecolor{myblue}{rgb}{0.2,0.2,0.8}
\definecolor{myzard}{cmyk}{0,0,0.05,0}
\definecolor{mywhite}{rgb}{1,1,1}
\definecolor{myred}{rgb}{1,0.,0.3}

\usepackage[colorlinks=true,citecolor=myblue,linkcolor=myred,urlcolor=myblue]{hyperref}

\def\be{\begin{equation}}
\def\ee{\end{equation}}
\def\ba{\begin{align}}
\def\enda{\end{align}}
\def\bi{\begin{itemize}}
\def\ei{\end{itemize}}

 \def\ee{\mathord{\rm e}}

 \def\ee{\mathord{\rm e}}

\renewcommand{\ee}{{\rm e}}

\def\beq{\begin{equation}}
\def\beq{\begin{equation}}
\def\eeq{\end{equation}}

 \newcommand{\ket}[1]{|#1\rangle}
 \newcommand{\bra}[1]{\langle #1|}

\begin{document}

\title[Short Title]{General scheme for the construction of a protected qubit subspace}

\author{N. Aharon\textsuperscript{1}, M. Drewsen\textsuperscript{2}, and A. Retzker\textsuperscript{3}}

\affiliation{\textsuperscript{1}School of Physics and Astronomy, Tel-Aviv University, Tel-Aviv 69978, Israel \\
 \textsuperscript{2}QUANTOP, Danish National Research Foundation Center for Quantum Optics, Department of Physics and Astronomy, University of Aarhus, DK-8000 {\AA}rhus C., Denmark \\
 \textsuperscript{3}Racah Institute of Physics, The Hebrew University of Jerusalem, Jerusalem 91904, Givat Ram, Israel
}

\date{\today}

\begin{abstract}
{We present a new robust decoupling scheme suitable for levels with either half integer or integer angular momentum states. Through continuous dynamical decoupling techniques, we create a protected qubit subspace, utilizing a multi-state qubit construction. Remarkably, the multi-state system can also be comprised of multiple sub-states within a single level. Our scheme can be realized with state-of-the-art experimental setups and thus has immediate applications for quantum information science. While the scheme is general and relevant for a multitude of solid state and atomic systems, we analyze its performance for the case composed of trapped ions. Explicitly, we show how single qubit gates and an ensemble coupling to a cavity mode  can be implemented efficiently. The scheme predicts a coherence time of $\sim1$ second, as compared to typically a few milliseconds for the bare states.}

\end{abstract}

\maketitle

{\em Introduction.---}
Protecting quantum bits (qubits) from decohernece due to interactions with their environment is a prime issue of experimental quantum information science. In the case of solid state and atomic qubits systems, the presence of ambient magnetic field fluctuations is in particular a problem. Consequently, several methods have been put forward to tackle this problem. The traditional solution is to utilize either a two-level sub-system of two integer total angular momentum states, which to first order has no Zeeman shifts \cite{Benhelm2008,Lucas2007,Olmschenk2007}, or a two-level system composed of two hyperfine states with identical first order shifts \cite{Langer2005,rare1}.
A third way is to use decoherence free subspaces \cite{Lidar1998,Kielpinski2001,Haeffner2005}, which requires spatially separated physical qubits to represent a single logic qubit and thus incurs considerable overhead, and is potentially vulnerable to decoherence due to field gradients.

Dynamical decoupling is another general strategy to tackle this problem \cite{Viola1998}. The pulsed version was proven to be extremely efficient \cite{Biercuk2009,Naydenov2011}; however, it may require complex pulse sequences.
The continuous version of dynamical decoupling \cite{Tan2013}  is based on spin locking \cite{Goldman}, where a continuous drive protects the system from external noise and weaker continuous pulses improves its robustness \cite{CaiCon}. Continuous dynamical decoupling could be combined in a natural way with gates \cite{NVGate} and could improve the coupling efficiency to superconducting cavities \cite{CaiCavity}.  However, both versions require composite schemes to overcome both the external (magnetic) noise and the controller (optical/microwave/rf) noise. A four level structure composed of the magnetic substates of two hyperfine levels with $F=0$ and $F=1$ has been designed to be perfectly robust to control fluctuations in conjunction with composite schemes \cite{Christof}, but this method is only applicable for this particular  spin system.

In this Letter we present a new and general method for the construction of a protected and robust qubit subspace. The method utilizes a multilevel structure, on which continuous dynamical decoupling fields are applied.  Our method is suitable for a wide range of solid state and atomic systems, and it is applicable to a variety of tasks in the field of quantum information science and quantum sensing, in particular, quantum magnetometery and quantum memories.  The method can be implemented with state of the art technology, and should be able to push the $T_2$ time to the $T_1$ limit.

{\em The general scheme.---} The general scheme defines the protected subspace which we denote by $\left\{  \left|D_{i}\right\rangle \right\} $. In the following $\mathrm{\mathbf{J}}$ is the angular momentum operator, $H_{d}$ is the (continuous) driving Hamiltonian, $\mathcal{H}_{D}$ is the Hilbert subspace of the protected (and hence dark) states and $\mathcal{H}_{\perp}$ is the complementary Hilbert space, that is, $\mathcal{H}=\mathcal{H}_{D} \oplus \mathcal{H}_{\perp}$. We define the protected subspace by
\begin{eqnarray}
\left\langle D_{j}\right|J_{z}\left|D_{i}\right\rangle =0 &  & \qquad \forall i,j,\nonumber \\
H_{d}\left|D_{i}\right\rangle =0 &  & \qquad \forall i.
\label{eq1}
\end{eqnarray}
The first equation ensures that the noise does not operate within the protected subspace; the noise can only cause transitions between a state in the protected subspace and a state in the complementary subspace. We assume (by construction) that for any eigenstate $\left|\psi_{i}\right\rangle \in \mathcal{H}_{\perp}$ of $H_d$ we have that $\vert \bra{\psi_{i}}H_{d}\ket{\psi_{i}} \vert$ is much larger than the characteristic frequency of the power spectrum of the noise \cite{norm_condition}. This ensures that the energy of all states in $\mathcal{H}_{D}$ is far from the energy of the states in $\mathcal{H}_{\perp}$ and thus the rate of  transitions from $\mathcal{H}_{D}$ to $\mathcal{H}_{\perp}$ due to noise is negligible. The second equation indicates that the protected subspace is the kernel of $H_d$ and hence, the protected states do not collect a dynamical phase and are immune to the noise originating from $H_d$. Note that these conditions are analogous to the error detection conditions in \cite{knill} since the errors are magnetic noise, which is represented by the $J_z$ operator, and fluctuations in $H_d$.

From the definition of the protected subspace we can also study the evolution within the subspace. Transitions between dark states can be generated by only one of the operators $J_{x}$ and $J_{y}$. Suppose that $J_{y}$ transforms between dark states, $J_{y}\left|D_{i}\right\rangle =\left|D_{j}\right\rangle $, $(i\neq j)$. Together with $J_{z}\left|D_{i}\right\rangle =\left|\varphi_{i}\right\rangle \in\mathcal{H}_{\perp}$ we have that $J_{y}J_{z}\left|D_{i}\right\rangle =\left|\tilde{\varphi}_{i}\right\rangle \in\mathcal{H}_{\perp}$
and $J_{z}J_{y}\left|D_{i}\right\rangle =\left|\varphi_{j}\right\rangle \in\mathcal{H}_{\perp}$, and hence $J_{x}\left|D_{i}\right\rangle \in\mathcal{H}_{\perp}$.
Whether it is $J_{x}$ or $J_{y}$ that transforms between the dark states is determined by $H_{d}$. Suppose again that $J_{y}$ transforms between the dark states. It is then easy to show that $[H_{d},J_{y}]\left|D_{i}\right\rangle=0$ and that $[H_{d},J_{x}]\left|D_{i}\right\rangle\in\mathcal{H}_{\perp}$. This limits the available \emph{direct} operations on the dark state to rotations around one axis. However, general unitary operations can be implemented by various methods \cite{Webster2013,Gatis2013,supp}. Since  $J_{z}\left|D_{i}\right\rangle =\left|\varphi_{i}\right\rangle \in\mathcal{H}_{\perp}$ we can also conclude that $\lceil \frac{d_{\mathcal{H}}}{2} \rceil \geq d_{\mathcal{H}_{D}}$ where the $d_{\mathcal{H}}$ and $ d_{\mathcal{H}_{D}}$ are the dimensions of the total Hilbert space and  the protected subspace respectively.

\begin{figure}
\centering
\includegraphics[width=1\columnwidth]{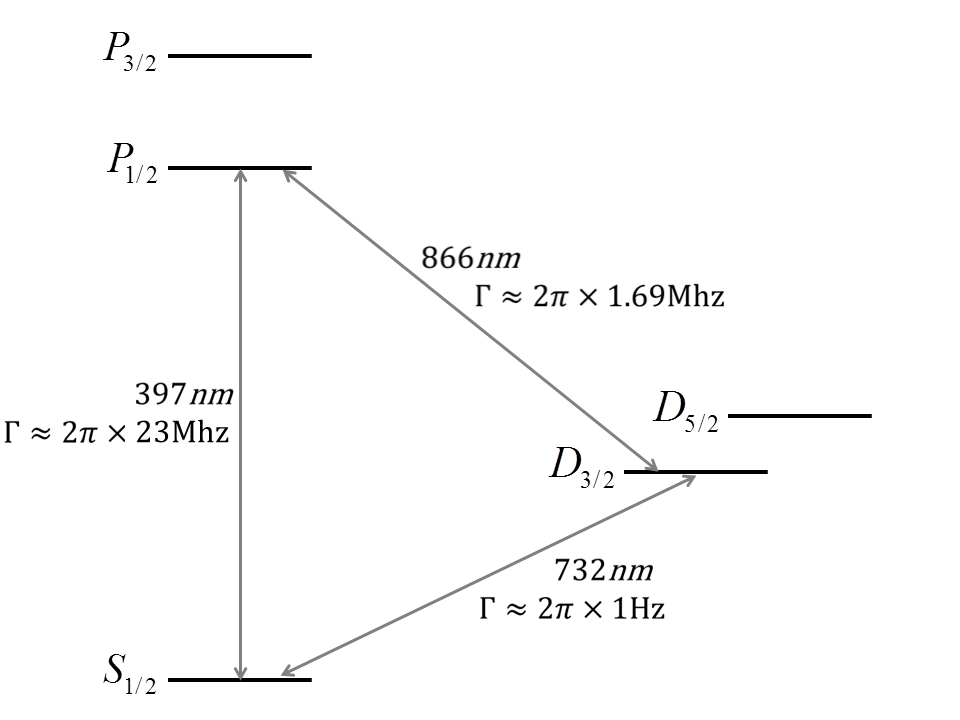}
\caption{ Level structure of the calcium ion, $^{40}\textrm{Ca}^{+}$. The $D_{3/2}$ subspace, which has a lifetime of $\sim1$ second, serves as the protected subspace. The $S_{1/2}-P_{1/2}$ transitions and the $D_{3/2}-P_{1/2}$ transitions are used in the initialization and construction of the protected subspace.}
\label{CalciumIon}
\end{figure}

{\em Implementation with trapped ions.---}
Below we present an implementation of the scheme with a system of trapped ions. Although the suggested implementation is applicable to a variety of ionic systems, we focus on the calcium ion (see Fig. \ref{CalciumIon}). Remarkably, the considered multi-state system is composed of multiple sub-states within a single level, specifically, the $D_{3/2}$ sub-levels. Since the $D_{3/2}$ states have a lifetime of $\sim1$ second, we consider their subspace to be the protected subspace. Please note that a very similar level structure exists for the barium ion with a longer lifetime of  $\sim20$ seconds.
For simplicity we will use the notation $\left|d_{3/2+m_{i}}\right\rangle\equiv\left|D_{3/2};m_{i}\right\rangle$, $\left|p_{1/2+m_{i}}\right\rangle\equiv\left|P_{1/2};m_{i}\right\rangle$ and $\left|s_{1/2+m_{i}}\right\rangle\equiv\left|S_{1/2};m_{i}\right\rangle$. The definition of the protected subspace given by Eq. (\ref{eq1}) results in the two dark states (see \cite{supp})

\begin{eqnarray}
\left|D_{1}\right\rangle  & = & \frac{\sqrt{3}}{2}\left|d_{1}\right\rangle -\frac{1}{2}\left|d_{3}\right\rangle, \nonumber \\
\left|D_{2}\right\rangle  & = & \frac{1}{2}\left|d_{0}\right\rangle - \frac{\sqrt{3}}{2}\left|d_{2}\right\rangle,
\label{eq2}
\end{eqnarray}
where it can be seen that the average magnetic moment for each state vanishes.

\begin{figure}
\centering
\includegraphics[width=1\columnwidth]{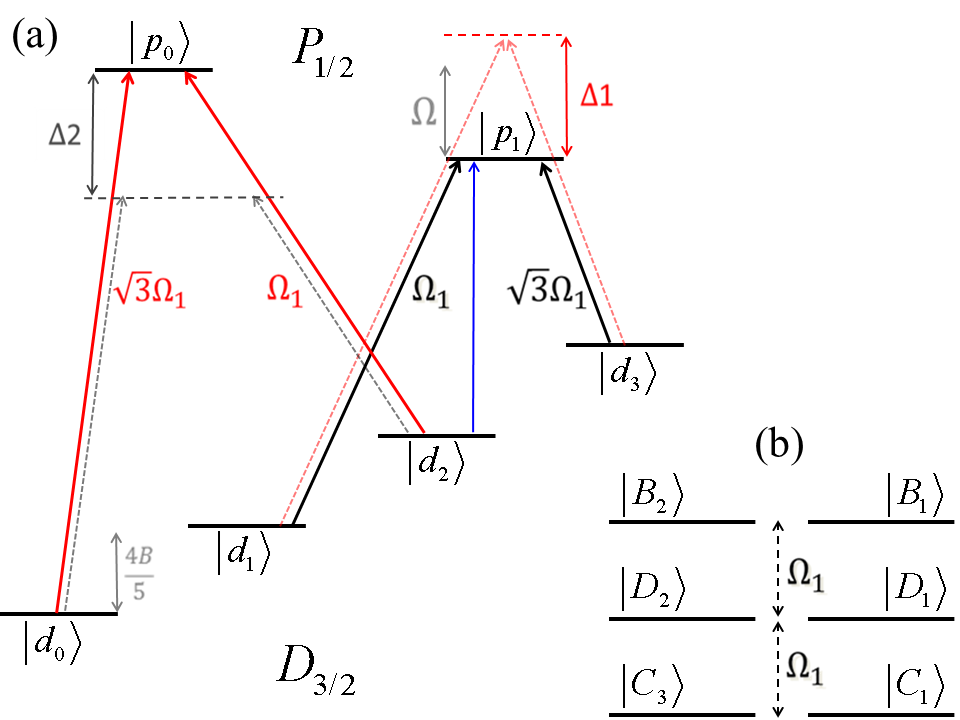}
\caption{ Realization of dark states. (a) The black (red) driving fields result in the first (second) dark state. The driving fields of each dark state also operate on the subspace of the other dark state (dashed lines), resulting in  small energy shifts. The detunings are given by $\Delta2=-\Delta1=\Omega+\frac{4B}{5}$, where $\Omega$ is the energy gap between the two $P_{1/2}$ states, introduced by the $S_{1/2}-P_{1/2}$ coupling. Blue arrow - optical pumping to the first dark state, $\left|D_{1}\right\rangle$. (b) Level structure in the dark states basis. The dark states $\left|D_{1}\right\rangle $ and $\left|D_{2}\right\rangle $ form the protected subspace. }
\label{Realization}
\end{figure}

These two orthonormal dark states can serve as a basis for a qubit memory.
The $D_{3/2}$ degeneracy is removed by applying a constant magnetic field along the $\hat{z}$ axes which results in an energy gap of $g_{J}B$  between any two adjacent energy levels, where $g_{J}=\frac{4}{5}$ is the Land\'{e} g-factor. A large enough $\left|B\right|$ such that $\left|g_{J}B\right|$ is much larger than the characteristic frequency of the noise, ensures that the dark states are also immune to $J_{x}$ and $J_{y}$ noise. We now describe the driving Hamiltonian, $H_d=H_{d1}+H_{d2}$. $H_{d1}$ corresponds to the simultaneous on-resonance coupling of the $\left|d_{1}\right\rangle $ and $\left|d_{3}\right\rangle $ states to the $\ket {p_{1}}$ state, and results in the first dark state, $\left|D_{1}\right\rangle $. $H_{d2}$ corresponds to the on-resonance coupling of the $\left|d_{0}\right\rangle $ and $\left|d_{2}\right\rangle $ states to the $\ket {p_{0}}$ state, and results in the second dark state, $\left|D_{2}\right\rangle $. However, the driving fields of each dark state can impact the other dark state since they operate on all of the $D_{3/2}$ states. We reduce this undesirable effect by creating an energy gap between the two $P_{1/2}$ states. This energy gap is achieved by the on-resonance coupling of the $\ket {s_{0}}$ and $\ket {p_{1}}$ states, and as a consequence, the driving fields of the first (second) dark state operate on the second (first) dark state with a detuning of $\Delta2=\Omega+\frac{4B}{5}$ ($\Delta1=-(\Omega+\frac{4B}{5})$) (see Fig. \ref{Realization}).

In the interaction picture and in the rotating wave approximation (RWA) the total driving Hamiltonian is given by

\begin{eqnarray}
  H_{d} &=& [(\frac{\Omega_{1}}{2}\left|p_{1}\right\rangle \left\langle d_{1}\right| + \frac{\sqrt{3}\Omega_{1}}{2}\left|p_{1}\right\rangle \left\langle d_{3}\right|)+h.c. \nonumber \\
   &+& (\frac{\Omega_{1}}{2}\left|p_{0}\right\rangle \left\langle d_{2}\right| +\frac{\sqrt{3}\Omega_{1}}{2}\left|p_{0}\right\rangle \left\langle d_{0}\right|)+h.c.  ],
\label{eq3}
\end{eqnarray}

This Hamiltonian has two eigenstates with zero eigenvalues, which are the desired dark states given by Eq. (\ref{eq2})
, and four bright eigenstates whose eigenvalues are equal to $\pm\Omega_1$ \cite{supp}.

So far we have discussed the construction of the protected subspace. In the following we estimate the lifetime, $T_1$, and the coherence time, $T_2$, of the dark states. The lifetime can be affected by the energy shifts caused by the driving fields of the other dark state, and the coherence time can be affected by the fluctuations of these energy shifts. The fluctuations in the energy shifts cause dephasing at a rate equal to the power spectrum of the noise at zero frequency. For the first dark state, $\left|D_{1}\right\rangle $, an energy shift fluctuation can also occur due to fluctuations of the driving field creating the energy gap between the two $P_{1/2}$ states.  Calculation of these energy shifts and their fluctuations (assuming a maximal fluctuation of $1\%$ in the intensity of the driving fields \cite{laser}), yields \cite{supp}

\begin{flushleft}
\begin{eqnarray}
\Delta E_1\leq\frac{\Omega_{1}^{2}}{4|\Delta1|}\biggl(1\pm\frac{3}{100}\biggr),\Delta E_2\leq\frac{\Omega_{1}^{2}}{4|\Delta2|}\biggl(1\pm\frac{2}{100}\biggr).\nonumber\\
\label{eq4}
\end{eqnarray}
\end{flushleft}

Both $\Delta E_1$ and $\Delta E_2$ are of the order of $\frac{\Omega_{1}^{2}}{\Omega}$, which for typical experimental setups is $\frac{\Omega_{1}^{2}}{\Omega}\sim\frac{\left(10^{5}\right)^{2}}{10^{9}}=10\mathrm{Hz}$. These energy shifts correspond to a small modification of the dark
states, $\left|D_{i}\right\rangle \rightarrow\sqrt{1-\epsilon}\left|D_{i}\right\rangle +\sqrt{\epsilon}\left|\varphi_{i}\right\rangle $,
where $\left|\varphi_{i}\right\rangle \in\mathcal{H}_{\perp}$, reducing the $T_1$ time from $1$ second to approximately $0.9$ seconds \cite{supp}.

The $T_2$ time can be affected by the fluctuations of $\Delta E1$ and $\Delta E2$. For the above experimental parameters we have that $T_2\leq\left(\Delta\left(\Delta E1-\Delta E2\right)\right)^{-1}\sim\left(\frac{\Omega_{1}^{2}}{100\Omega}\right)^{-1}=10\sec$ \cite{T2_limit,supp}. As this bound is even larger than $T_1$, we conclude that the fluctuations in the driving fields do not reduce the $T_2$ time.
In addition, relative amplitude and phase fluctuations will limit the  $T_2$ time by  $\frac{T_2^*}{\eta^2}$, where $T_2^*$ is the coherence time of the bare states, and $\eta$ is the rate of the relative amplitude fluctuations; since these are usually small we can neglect this correction.

Another source of noise comes from polarization imperfections. The typical experimental error in the polarization is $\sim1\%$. This means that $\sim1\%$ of a $\sigma^{+}$ polarized beam is actually $\sigma^{-}$ polarized and vice versa, causing an error within the driving of each dark state (for example, $1\%$ of the $\sigma^{-}$ beam which couples the  $\ket {d_{3}}$ and $\ket {p_{1}}$ states is a $\sigma^{+}$ beam which couples the $\ket {d_{1}}$ and $\ket {p_{1}}$ states). The polarization errors also cause an energy shift and modify the dark state. However, an energy gap of $\sim10\textrm{Mhz}$ between the $D_{3/2}$ states (due to the zeeman splitting) ensures that neither the $T_1$ time or the $T_2$ time are reduced \cite{supp}.

We have thus constructed a protected and robust qubit subspace with a lifetime and a coherence time which are almost identical to the $D_{3/2}$ lifetime, equaling approximately $0.9$ seconds, while the $T_2^*$ time is of the order of $1\textrm{ms}$ \cite{Herskind2009,Albert2011}.

{\em Initialization and single qubit gates.---}
By adding two extra laser beams, one that couples the $\ket {s_{1}}$ state to the $\ket {p_{1}}$ state, and the other that couples the $\ket {d_{2}}$ state to the $\ket {p_{1}}$ state (blues laser in Fig. \ref{Realization}), we can achieve optical pumping to the dark states $\ket {D_{1}}$. This way, the dark state $\ket {D_{2}}$ is taken out of the protected subspace, but because of $H_d$ the state will eventually evolve to the dark state $\ket {D_{1}}$.  Another method of initialization is to optically pump into the $\ket {d_{3}}$ state, and then conduct a STIRAP procedure via a Raman transition.

\begin{figure}
\centering
\includegraphics[scale=0.37]{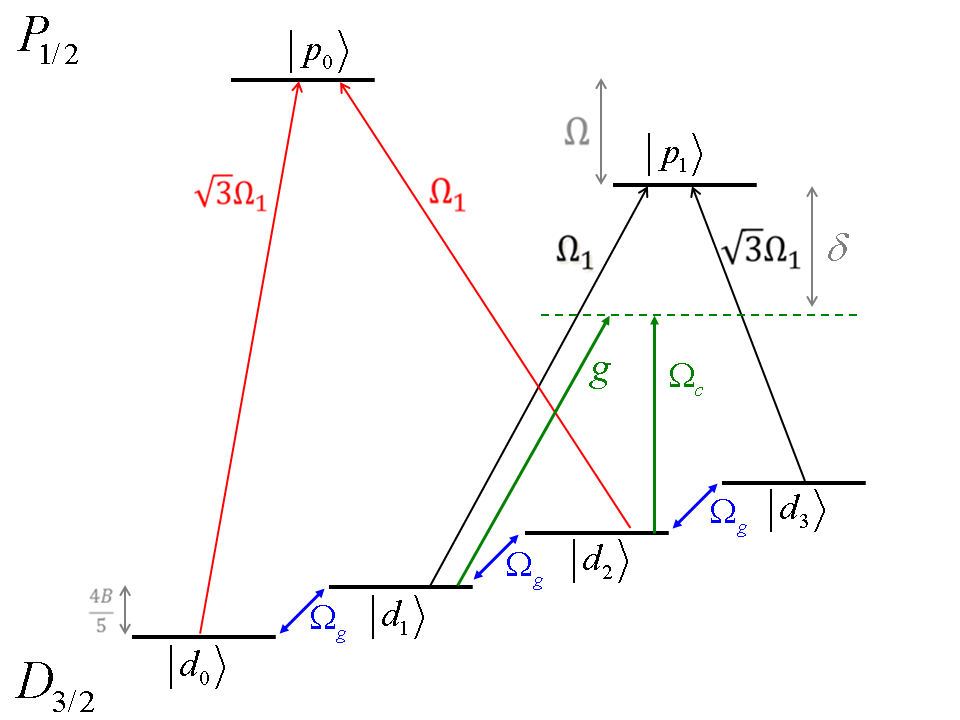} 
\caption{ Realization of (i) a single qubit gate (blue) (ii) coupling to a cavity mode (green). }
\label{Gates}
\end{figure}

We propose an experimentally simple method for the implementation of a single qubit $\sigma_{y}$ gate by  applying a microwave field which is set to be on-resonance with the energy gap between the $D_{3/2}$ states (see Fig. \ref{Gates}). More specifically, the microwave field is tuned to apply the $J_{y}$ operator, as in our case $[H_{d},J_{y}]\ket {D_{i}}=0$.
In the interaction picture and in the RWA the Hamiltonian of the single qubit gate is given by
\begin{eqnarray}
H_{g}=i\Omega_{g}\left(\frac{\sqrt{3}}{2}\left|d_{1}\right\rangle \left\langle d_{0}\right|+\left|d_{2}\right\rangle \left\langle d_{1}\right|+\frac{\sqrt{3}}{2}\left|d_{3}\right\rangle \left\langle d_{2}\right|\right)+h.c.,\nonumber\\
\end{eqnarray}
which corresponds to the operator $-\frac{i3\Omega_{g}}{2}\left|D_{2}\right\rangle \left\langle D_{1}\right|$ in the dark states basis \cite{supp}. In the Supplementary Material we explicitly show how to construct $\sigma_{x}$ and $\sigma_{z}$ gates, which allow for the implementation of any single qubit unitary operation \cite{supp}.

{\em Interaction with a cavity mode.---}
One of the most important applications of robust quantum states is the implementation of a quantum memory. For this purpose, it is also necessary to have an efficient interaction between the robust states of the quantum memory and the mediating system which delivers the data to be stored and retrieved from memory. Here, we focus on the interaction of ions with a cavity mode, as several experimental investigations are currently exploring this situation \cite{Keller2007,Herskind2009,Dubin2010,Albert2011,Stute2012,Stute2013}. Such an interaction will not only allow for the implementation of a quantum memory but could also allow for multi-qubit gates where the interaction between different qubits is mediated via the cavity modes.

We begin by setting the cavity mode such that its frequency and polarization corresponds to the detuned coupling of the $\left|d_{1}\right\rangle $ state to the $\ket {p_{1}}$ state with the detuning $\delta$ to be specified below.
In addition, we apply an external control field which corresponds to the detuned
coupling of the $\left|d_{2}\right\rangle $ state to the $\ket {p_{1}}$
state with the same detuning $\delta$ and with a Rabi frequency $\Omega_{c}$ such that $\delta\gg\Omega_{c}\gg g$, where $g$ is the rate describing the coupling between a single photon in the cavity mode to a single ion (see Fig. \ref{Gates}). This interaction
couples the $\left|d_{1}\right\rangle $ and $\left|d_{2}\right\rangle $
states and results in the effective Hamiltonian $H_{eff}=-\frac{g\Omega_{c}}{2\delta}\left(\left|d_{2}\right\rangle \left\langle d_{1}\right|a+h.c.\right)$, where $a$ is the annihilation operator of the cavity mode.
In the dark states basis the interaction, which is given by $H_{eff}\approx-\frac{3g\Omega_{c}}{8\delta}\left(\left|D_{2}\right\rangle \left\langle D_{1}\right|a+h.c.\right)$,
couples a cavity mode to a robust qubit \cite{supp}.
However, the strength of this coupling is usually weak compared to the cavity and an ion damping rates. That
is, $\kappa,\gamma\gg\frac{3g\Omega_{c}}{8\delta}$, where $\kappa$ is the cavity's damping rate and
and $\gamma=\frac{\Omega_{c}^{2}}{\delta^{2}} \Gamma_{p_{1}}$ is the ion's damping rate. This is known as the weak coupling regime in which transmission of quantum information
is not possible. The problem can be circumvented by coupling a cavity
mode to an ensemble of ions. The coupling strength is enhanced by
$\sqrt{N}$, where $N$ is the effective number of ions, and for a large enough ensemble this results in $\kappa,\gamma\ll\sqrt{N}\frac{3g\Omega_{c}}{8\delta}$, which are the conditions for the collective strong coupling regime. (Note that since we consider $N$ ions, the probability of emitting a photon is $\sim\frac{\Omega_{c}^{2}}{\delta^{2}} N$. However, the factor $N$ is canceled out because the interaction results in a Dicke state). From the condition of the strong coupling regime on the ion's damping rate we must have that $\frac{\Omega_{c}}{\delta}\ll \frac{3 g \sqrt{N}}{8 \Gamma_{p_{1}}}$. Substituting
$\Gamma_{p_{1}} = 2\pi \times  23\textrm{MHz}$, $g = 2\pi \times  0.5\textrm{MHz}$, and $\sqrt{N}\sim10$ (which could be achieved e.g. as a string of one species of ions within another species \cite{galaxy}), we get that $\frac{\Omega_{c}}{\delta}\ll \frac{1}{10}$, and thus we set  $\frac{\Omega_{c}}{\delta}\sim10^{-2}$. The condition on the cavity's damping rate then implies that $\kappa\ll\frac{g}{10}\sim \pi \times  0.1\textrm{MHz}$ is required. Such damping rates are exhibited in current high-finesse cavities.

Note that by removing the control field we are left only with the coupling
to the cavity mode which results in the Hamiltonian $H_{R}=-\frac{g^{2}}{\delta}\left|d_{1}\right\rangle \left\langle d_{1}\right|a^{\dagger}a$,
corresponding to $H_{R}\approx-\frac{3g^{2}}{4\delta}\left|D_{1}\right\rangle \left\langle D_{1}\right|a^{\dagger}a$
in the dark states basis. As $H_{R}$ takes $\frac{1}{\sqrt{2}}\left(\left|D_{2}\right\rangle +\left|D_{1}\right\rangle \right)$
to $\frac{1}{\sqrt{2}}\left(\left|D_{2}\right\rangle +e^{i\frac{3 g^{2}t}{4\delta}a^{\dagger}a}\left|D_{1}\right\rangle \right)$
a non-demolition measurement of the photon-number in the cavity can be done by a Ramsey spectroscopy experiment on the dark states \cite{supp}. This constitutes an alternative strategy to electron shelving based methods \cite{Clausen2013}.

{\em Discussion.---}
A scheme for robust qubits based on continuous dynamical decoupling was presented.
The scheme is general in the sense that it can be applied to all systems satisfying  Eq. (\ref{eq1}), but in addition can have different characteristics. Unlike most commonly used methods, our scheme is applicable to systems with half integer total angular momentum.

Although our example utilizes the $D_{3/2}$ subspace, in principle, the scheme can also be applied to subspaces of a different total angular momentum, such as the $D_{5/2}$ subspace of the calcium ion. In this case, a protected qubit subspace can be achieved by first, an on-resonance $J_x$ coupling of all $P_{3/2}$ states (which results in four $J_x$ eigenstates) , and second, by the on resonance coupling of the $\ket {d_{0}}$ and $\ket {d_{5}}$ states to one of the above eigenstates (resulting in one dark state), and by the on resonance coupling of the $\ket {d_{2}}$ and $\ket {d_{3}}$ states to another $J_x$ eigenstate (resulting in a second dark state). The ability to couple negative angular momentum states with positive angular momentum states constitutes a necessary condition for satisfying  Eq. (\ref{eq1}).

The scheme was analyzed in detail for a system of trapped ions based on optical control, in which the quantum memory consists of a string of ions that could either exist on its own, or inside a larger crystal of a different species \cite{galaxy}. The simplicity of the scheme, which does not require complex laser pulses, enlarges the scope of quantum memories to laser control, and provides new perspectives for laser manipulations.

By combining the setup with a stripline resonator, a conversion between an optical photon to a microwave photon could be achieved. Our scheme can also be realized with barium ions which have a lifetime of $\sim20$ seconds. Such a long lifetime would enable a relaxation of the requirements on the number of ions and the cavity damping rate, resulting in a simpler experimental realization, and would also increase the storage time by one further order of magnitude.

{\it Acknowledgements.---} We acknowledge support from the Israel Science Foundation (Grant No. 1125/10) and the Binational Science Foundation (Grant No. 32/08) (N.A), the Carlsberg Foundation and the EU via the FP7 projects 'Physics of Ion Coulomb Crystals' (PICC) and 'Circuit and Cavity Quantum Electrodynamics' (CCQED) (M.D), and a carrier integration grant(CIG) no. 321798 IonQuanSense FP7-PEOPLE-2012-CIG (A.R).

\newpage

\begin{widetext}
\section{Supplementary material}

In what follows we present detailed derivations of the results concerning
the implementation of the scheme with a system of trapped ions.

\section{Construction of dark states}

We consider the subspace of the $D_{3/2}$ level of the calcium ion
as a basis for the protected subspace. Under the Zeeman splitting,
we assume that the noise operating on the system is generally given
by $f(t)J_{z}$, where $f(t)$ is some random function of time. Hence,
the Hamiltonian describing the noise is given by

\begin{equation}
H_{noise}=f(t)\left(\frac{3}{2}\left|d_{3}\right\rangle \left\langle d_{3}\right|+\frac{1}{2}\left|d_{2}\right\rangle \left\langle d_{2}\right|-\frac{1}{2}\left|d_{1}\right\rangle \left\langle d_{1}\right|-\frac{3}{2}\left|d_{0}\right\rangle \left\langle d_{0}\right|\right).
\end{equation}

Define the dark states $\left|D_{1}\right\rangle =\alpha\left|d_{1}\right\rangle +\beta\left|d_{3}\right\rangle $
and $\left|D_{2}\right\rangle =\gamma\left|d_{0}\right\rangle +\delta\left|d_{2}\right\rangle $,
where $\left|\alpha\right|^{2}+\left|\beta\right|^{2}=\left|\gamma\right|^{2}+\left|\delta\right|^{2}=1$,
and their orthogonal bright states, $\left|D_{1}^{\perp}\right\rangle =\beta^{*}\left|d_{1}\right\rangle -\alpha^{*}\left|d_{3}\right\rangle $
and $\left|D_{2}^{\perp}\right\rangle =\delta^{*}\left|d_{0}\right\rangle -\gamma^{*}\left|d_{2}\right\rangle $.
The definition of the protected subspace (Eq.(1) in the main text)
implies that $H_{noise}\left|D_{i}\right\rangle =0$ and $H_{d}\left|D_{i}\right\rangle =0$,
where $H_{d}$ is the continuous driving Hamiltonian. The first requirement
suggests that
\begin{eqnarray}
H_{noise}\left|D_{1}\right\rangle = & f(t)\left(\frac{3}{2}\left|d_{3}\right\rangle \left\langle d_{3}\right|-\frac{1}{2}\left|d_{1}\right\rangle \left\langle d_{1}\right|\right)\left|D_{1}\right\rangle \nonumber \\
\approx & f(t)\left(\left(\frac{3}{2}\left|\beta\right|^{2}-\frac{1}{2}\left|\alpha\right|^{2}\right)\left|D_{1}\right\rangle \left\langle D_{1}\right|+\left(\frac{3}{2}\left|\alpha\right|^{2}-\frac{1}{2}\left|\beta\right|^{2}\right)\left|B_{1}\right\rangle \left\langle B_{1}\right|+\ldots\right)\left|D_{1}\right\rangle =0,
\end{eqnarray}
and hence, we must have that $\left|\alpha\right|^{2}=3\left|\beta\right|^{2}$.
Recall that we assume (by construction) that the rate of transitions
from $\mathcal{H}_{D}$ to $\mathcal{H}_{\perp}$ due to noise is
negligible , and therefore we neglect the off-diagonal terms. Similarly,

\begin{eqnarray}
H_{noise}\left|D_{2}\right\rangle = & f(t)\left(\frac{1}{2}\left|d_{2}\right\rangle \left\langle d_{2}\right|-\frac{3}{2}\left|d_{0}\right\rangle \left\langle d_{0}\right|\right)\left|D_{2}\right\rangle \nonumber \\
\approx & f(t)\left(\left(\frac{3}{2}\left|\gamma\right|^{2}-\frac{1}{2}\left|\delta\right|^{2}\right)\left|D_{2}\right\rangle \left\langle D_{2}\right|+\left(\frac{3}{2}\left|\delta\right|^{2}-\frac{1}{2}\left|\gamma\right|^{2}\right)\left|B_{2}\right\rangle \left\langle B_{2}\right|+\ldots\right)\left|D_{2}\right\rangle =0,
\end{eqnarray}
 and hence, we must have that $\left|\delta\right|^{2}=3\left|\gamma\right|^{2}$.

The relative phases of the dark states are now determined by the second
requirement, $H_{d}\left|D_{i}\right\rangle =0$. In our
example $H_{d}=H_{d1}+H_{d2}$, where $H_{d1}$ and $H_{d2}$ are
generally given by (in the interaction picture (IP) and in the rotating
wave approximation (RWA), as will be derived below)
\begin{eqnarray}
H_{d1} & = & \frac{\Omega_{1}}{2}\left|p_{1}\right\rangle \left\langle d_{1}\right|+\frac{\Omega_{2}}{2}\left|p_{1}\right\rangle \left\langle d_{3}\right|+h.c.,\nonumber \\
H_{d2} & = & \frac{\Omega_{3}}{2}\left|p_{0}\right\rangle \left\langle d_{2}\right|+\frac{\Omega_{4}}{2}\left|p_{0}\right\rangle \left\langle d_{0}\right|+h.c..
\end{eqnarray}
From the second requirement, $H_{d}\left|D_{i}\right\rangle =0$,
it then follows that
\begin{equation}
\frac{\Omega_{2}}{\Omega_{1}}=-\frac{\alpha}{\beta}=\sqrt{3},
\end{equation}

and
\begin{equation}
\frac{\Omega_{4}}{\Omega_{3}}=-\frac{\delta}{\gamma}=\sqrt{3}.
\end{equation}

We conclude that the dark states, which satisfy the requirements of
a protected qubit subspace are

\begin{eqnarray}
\left|D_{1}\right\rangle  & = & \frac{\sqrt{3}}{2}\left|d_{1}\right\rangle -\frac{1}{2}\left|d_{3}\right\rangle ,\nonumber \\
\left|D_{2}\right\rangle  & = & \frac{1}{2}\left|d_{0}\right\rangle -\frac{\sqrt{3}}{2}\left|d_{2}\right\rangle .
\end{eqnarray}

Setting $\Omega_{3}=\Omega_{1}$, we have that
\begin{equation}
H_{d}=\left(\frac{\Omega_{1}}{2}\left|p_{1}\right\rangle \left\langle d_{1}\right|+\frac{\sqrt{3}\Omega_{1}}{2}\left|p_{1}\right\rangle \left\langle d_{3}\right|+\frac{\Omega_{1}}{2}\left|p_{0}\right\rangle \left\langle d_{2}\right|+\frac{\sqrt{3}\Omega_{1}}{2}\left|p_{0}\right\rangle \left\langle d_{0}\right|\right)+h.c..
\label{HdRWA}
\end{equation}

Denote by $\Delta$,$B$, and $\Omega$ the energy gap between the
$\left|d_{3}\right\rangle $ and $\left|p_{1}\right\rangle $ states,
the amplitude of the static magnetic field, and the energy gap between
the $\left|p_{0}\right\rangle $ and $\left|p_{1}\right\rangle $
states (which is introduced by an on-resonance coupling of the $\left|s_{0}\right\rangle $
and $\left|p_{1}\right\rangle $ states) respectively, and note that
the Landé g-factor is $g_{J}=\frac{4}{5}$. We then have that

\begin{eqnarray}
H_{d1}^{0} & = & \Delta\left|p_{1}\right\rangle \left\langle p_{1}\right|-\frac{8B}{5}\left|d_{1}\right\rangle \left\langle d_{1}\right|,\nonumber \\
H_{d2}^{0} & = & \left(\Delta+\Omega\right)\left|p_{0}\right\rangle \left\langle p_{0}\right|-\frac{12B}{5}\left|d_{0}\right\rangle \left\langle d_{0}\right|-\frac{4B}{5}\left|d_{2}\right\rangle \left\langle d_{2}\right|,
\end{eqnarray}
 and hence,

\begin{eqnarray}
H_{d1} & = & H_{d1}^{0}+\Omega_{1}\cos\left[\left(\Delta+\frac{8B}{5}\right)t\right]\left(\left|p_{1}\right\rangle \left\langle d_{1}\right|+h.c.\right)+\sqrt{3}\Omega_{1}\cos\left[\Delta t\right]\left(\left|p_{1}\right\rangle \left\langle d_{3}\right|+h.c.\right),\nonumber \\
H_{d2} & = & H_{d2}^{0}+\Omega_{1}\cos\left[\left(\Delta+\Omega+\frac{4B}{5}\right)t\right]\left(\left|p_{0}\right\rangle \left\langle d_{2}\right|+h.c.\right)+\sqrt{3}\Omega_{1}\cos\left[\left(\Delta+\Omega+\frac{12B}{5}\right)t\right]\left(\left|p_{0}\right\rangle \left\langle d_{2}\right|+h.c.\right).\nonumber \\
\end{eqnarray}
Moving to the IP with respect to $H_{d1}^{0}$ and $H_{d2}^{0}$,
and making the RWA we arrive at Eq. (\ref{HdRWA}),

\begin{eqnarray}
H_{d} & = & \left(\frac{\Omega_{1}}{2}\left|p_{1}\right\rangle \left\langle d_{1}\right|+\frac{\sqrt{3}\Omega_{1}}{2}\left|p_{1}\right\rangle \left\langle d_{3}\right|+\frac{\Omega_{1}}{2}\left|p_{0}\right\rangle \left\langle d_{2}\right|+\frac{\sqrt{3}\Omega_{1}}{2}\left|p_{0}\right\rangle \left\langle d_{0}\right|\right)+h.c.\nonumber \\
 & = & 0\left(\left|D_{1}\right\rangle \left\langle D_{1}\right|+\left|D_{2}\right\rangle \left\langle D_{2}\right|\right)+\Omega_{1}\left(\left|B_{1}\right\rangle \left\langle B_{1}\right|+\left|B_{2}\right\rangle \left\langle B_{2}\right|\right)-\Omega_{1}\left(\left|C_{1}\right\rangle \left\langle C_{1}\right|+\left|C_{2}\right\rangle \left\langle C_{2}\right|\right),
\end{eqnarray}
 where
\begin{eqnarray*}
\left|D_{1}\right\rangle  & = & \frac{\sqrt{3}}{2}\left|d_{1}\right\rangle -\frac{1}{2}\left|d_{3}\right\rangle ,\\
\left|B_{1}\right\rangle  & = & \frac{1}{2\sqrt{2}}\left|d_{1}\right\rangle +\frac{\sqrt{3}}{2\sqrt{2}}\left|d_{3}\right\rangle +\frac{1}{\sqrt{2}}\left|p_{1}\right\rangle ,\\
\left|C_{1}\right\rangle  & = & -\frac{1}{2\sqrt{2}}\left|d_{1}\right\rangle -\frac{\sqrt{3}}{2\sqrt{2}}\left|d_{3}\right\rangle +\frac{1}{\sqrt{2}}\left|p_{1}\right\rangle ,
\end{eqnarray*}
 and

\begin{eqnarray*}
\left|D_{2}\right\rangle  & = & \frac{1}{2}\left|d_{0}\right\rangle -\frac{\sqrt{3}}{2}\left|d_{2}\right\rangle ,\\
\left|B_{2}\right\rangle  & = & \frac{\sqrt{3}}{2\sqrt{2}}\left|d_{0}\right\rangle +\frac{1}{2\sqrt{2}}\left|d_{2}\right\rangle +\frac{1}{\sqrt{2}}\left|p_{0}\right\rangle ,\\
\left|C_{2}\right\rangle  & =- & \frac{\sqrt{3}}{2\sqrt{2}}\left|d_{0}\right\rangle -\frac{1}{2\sqrt{2}}\left|d_{2}\right\rangle +\frac{1}{\sqrt{2}}\left|p_{0}\right\rangle .
\end{eqnarray*}

Note that the unitary transformations which takes from the $\left\{ \left|d_{1}\right\rangle ,\left|d_{3}\right\rangle ,\left|p_{1}\right\rangle \right\} $
basis to the $\left\{ \left|D_{1}\right\rangle ,\left|B_{1}\right\rangle ,\left|C_{1}\right\rangle \right\} $
basis, and from the $\left\{ \left|d_{0}\right\rangle ,\left|d_{2}\right\rangle ,\left|p_{0}\right\rangle \right\} $
basis to the $\left\{ \left|D_{2}\right\rangle ,\left|B_{2}\right\rangle ,\left|C_{2}\right\rangle \right\} $
basis are given
\begin{equation}
U_{D1}=\left(\begin{array}{ccc}
\frac{\sqrt{3}}{2} & -\frac{1}{2} & 0\\
\frac{1}{2\sqrt{2}} & \frac{\sqrt{3}}{2\sqrt{2}} & \frac{1}{\sqrt{2}}\\
-\frac{1}{2\sqrt{2}} & -\frac{\sqrt{3}}{2\sqrt{2}} & \frac{1}{\sqrt{2}}
\end{array}\right),\quad\textrm{and}\quad U_{D2}=\left(\begin{array}{ccc}
\frac{1}{2} & -\frac{\sqrt{3}}{2} & 0\\
\frac{\sqrt{3}}{2\sqrt{2}} & \frac{1}{2\sqrt{2}} & \frac{1}{\sqrt{2}}\\
-\frac{\sqrt{3}}{2\sqrt{2}} & -\frac{1}{2\sqrt{2}} & \frac{1}{\sqrt{2}}
\end{array}\right)
\end{equation}
 respectively.

\section{Estimation of the $T_{1}$ and $T_{2}$ times}

\subsection{The effect of all driving fields and their fluctuations}

In this section we derive estimations for the lifetime $T_{1}$
and the coherence time $T_{2}$ of the dark states. Since the driving
fields operate on all of the $D_{3/2}$ states, the driving fields
of each dark state also operate on the sub-levels of the other dark
state. For this reason we introduce the energy gap between the two
$P_{1/2}$ states. However, the driving fields of each dark state
still cause a small energy shift to the other dark state and slightly
modify it, which may reduce the $T_{1}$ time. Fluctuations in these
energy shifts, caused by intensity fluctuations of the driving fields,
may also reduce the $T_{2}$ time. In addition, for the first dark
state $\left|D_{1}\right\rangle $ energy fluctuations also occur
due to intensity fluctuations of the $\left|s_{0}\right\rangle $
and $\left|p_{1}\right\rangle $ coupling field. In order to obtain
an upper bound of $T_{1}$ and $T_{2}$, we take the worst case scenario
of a $1\%$ fluctuation of all driving fields \cite{laser}. This leads to

\begin{eqnarray}
H_{D1} & = & H_{d1}\pm\frac{\Omega}{100}\left|p_{1}\right\rangle \left\langle p_{1}\right|\nonumber \\
 & + & \sqrt{3}\left(\Omega_{1}\pm\frac{\Omega_{1}}{100}\right)\cos\left[\left(\Delta+\frac{8B}{5}-\Delta1\right)t\right]\left(\left|p_{1}\right\rangle \left\langle d_{1}\right|+h.c.\right)\nonumber \\
 & + & \left(\Omega_{1}\pm\frac{\Omega_{1}}{100}\right)\cos\left[\left(\Delta-\Delta1\right)t\right]\left(\left|p_{1}\right\rangle \left\langle d_{3}\right|+h.c.\right),
\end{eqnarray}

and
\begin{eqnarray}
H_{D2} & = & H_{d2}\nonumber \\
 & + & \sqrt{3}\left(\Omega_{1}\pm\frac{\Omega_{1}}{100}\right)\cos\left[\left(\Delta+\Omega+\frac{4B}{5}-\Delta2\right)t\right]\left(\left|p_{0}\right\rangle \left\langle d_{2}\right|+h.c.\right)\nonumber \\
 & + & \left(\Omega_{1}\pm\frac{\Omega_{1}}{100}\right)\cos\left[\left(\Delta+\Omega+\frac{12B}{5}-\Delta2\right)t\right]\left(\left|p_{0}\right\rangle \left\langle d_{2}\right|+h.c.\right).
\end{eqnarray}
where $\Delta1$ and $\Delta2$ are the detunings (see Fig. 2 in main
text). We assume that the two driving fields of each dark state have
the same source, and thus, intensity fluctuations of these driving
fields do not cause an energy shift to that dark state and do not
modify it. Moving to the dark states basis in the IP and taking the RWA, we have
that

\begin{eqnarray}
H_{D1} & = & 0\left|D_{1}\right\rangle \left\langle D_{1}\right|+\left(\Omega_{1}\pm\frac{\Omega}{200}\right)\left|B_{1}\right\rangle \left\langle B_{1}\right|-\left(\Omega_{1}\mp\frac{\Omega}{200}\right)\left|C_{1}\right\rangle \left\langle C_{1}\right|\nonumber \\
 & + & \frac{e^{-i\Delta1t}\left(\Omega_{1}\pm\frac{\Omega_{1}}{100}\right)}{2\sqrt{2}}\left(\left|B_{1}\right\rangle \left\langle D_{1}\right|+\left|C_{1}\right\rangle \left\langle D_{1}\right|\right)+h.c.\nonumber \\
 & \pm & \frac{\Omega}{200}\left|B_{1}\right\rangle \left\langle C_{1}\right|+h.c.\,,
\end{eqnarray}
and
\begin{eqnarray}
H_{D2} & = & 0\left|D_{2}\right\rangle \left\langle D_{2}\right|+\Omega_{1}\left|B_{2}\right\rangle \left\langle B_{2}\right|-\Omega_{1}\left|C_{2}\right\rangle \left\langle C_{2}\right|\nonumber \\
 & - & \frac{e^{-i\Delta2t}\left(\Omega_{1}\pm\frac{\Omega_{1}}{100}\right)}{2\sqrt{2}}\left(\left|B_{2}\right\rangle \left\langle D_{2}\right|+\left|C_{2}\right\rangle \left\langle D_{2}\right|\right)+h.c.\,.
\end{eqnarray}
We now move again to the IP with respect to $H_{I1}^{0}=\Delta1\left|D_{1}\right\rangle \left\langle D_{1}\right|$
and $H_{I2}^{0}=\Delta2\left|D_{2}\right\rangle \left\langle D_{2}\right|$
and obtain

\begin{eqnarray}
H_{D1} & = & -\Delta1\left|D_{1}\right\rangle \left\langle D_{1}\right|+\left(\Omega_{1}\pm\frac{\Omega}{200}\right)\left|B_{1}\right\rangle \left\langle B_{1}\right|-\left(\Omega_{1}\mp\frac{\Omega}{200}\right)\left|C_{1}\right\rangle \left\langle C_{1}\right|\nonumber \\
 & + & \frac{\left(\Omega_{1}\pm\frac{\Omega_{1}}{100}\right)}{2\sqrt{2}}\left(\left|B_{1}\right\rangle \left\langle D_{1}\right|+\left|C_{1}\right\rangle \left\langle D_{1}\right|\right)+h.c.\nonumber \\
 & \pm & \frac{\Omega}{200}\left|B_{1}\right\rangle \left\langle C_{1}\right|+h.c.\,,
\end{eqnarray}
and
\begin{eqnarray}
H_{D2} & = & -\Delta2\left|D_{2}\right\rangle \left\langle D_{2}\right|+\Omega_{1}\left|B_{2}\right\rangle \left\langle B_{2}\right|-\Omega_{1}\left|C_{2}\right\rangle \left\langle C_{2}\right|\nonumber \\
 & - & \frac{\left(\Omega_{1}\pm\frac{\Omega_{1}}{100}\right)}{2\sqrt{2}}\left(\left|B_{2}\right\rangle \left\langle D_{2}\right|+\left|C_{2}\right\rangle \left\langle D_{2}\right|\right)+h.c.\,.
\end{eqnarray}
In first order of $\frac{\Omega_{1}^{2}}{\left|\Delta1\right|}$ and
$\frac{\Omega_{1}^{2}}{\left|\Delta2\right|}$, the coupling of the
dark states to the $\left|B_{i}\right\rangle $ and $\left|C_{i}\right\rangle $
states results in the (maximal) energy shifts

\begin{equation}
\Delta E_{1}=\frac{\Omega_{1}^{2}}{4\left|\Delta1\right|}\,,\quad\Delta E_{2}=\frac{\Omega_{1}^{2}}{4\left|\Delta2\right|}\,,
\end{equation}
and their (maximal) fluctuations
\begin{equation}
\Delta\left(\Delta E_{1}\right)=\frac{2}{400}\frac{\Omega_{1}^{2}}{\left|\Delta1\right|}+\frac{\Omega_{1}^{2}}{4\left|\Delta1\right|}\frac{\Omega}{100\left|\Delta1\right|}\approx\frac{3}{400}\frac{\Omega_{1}^{2}}{\left|\Delta1\right|}\,,\quad\Delta\left(\Delta E_{2}\right)=\frac{2}{400}\frac{\Omega_{1}^{2}}{\left|\Delta2\right|}\,.
\end{equation}
The energy shifts correspond to a small modification of the dark states,
taking $\left|D_{i}\right\rangle $ to $\sqrt{1-\epsilon}\left|D_{i}\right\rangle +\sqrt{\epsilon}\left|\varphi_{i}\right\rangle $,
where $\left|\varphi_{i}\right\rangle \in\mathcal{H}_{\perp}$. In
the first order of $\left(\frac{\Omega_{1}}{\Delta1}\right)^{2}$
and $\left(\frac{\Omega_{1}}{\Delta2}\right)^{2}$ we find that $\epsilon_{i}=\frac{1}{2}\left(\frac{\Omega_{1}}{\Delta i}\right)^{2}$.
Taking into account that the probability to be in a $P_{1/2}$ state
given that the ion is in a $\left|B_{i}\right\rangle $ or a$\left|C_{i}\right\rangle $
state is $\frac{1}{2}$, and setting $\Omega_{1}\sim10^{5}\,\textrm{Hz}$
and $\Omega\sim10^{9}\,\textrm{Hz}$, the lifetime is given by
\begin{equation}
T_{1}=\frac{1}{P\left(\left|p_{i}\right\rangle \right)\Gamma_{P_{1/2}}+P\left(\left|d_{i}\right\rangle \right)\Gamma_{D_{3/2}}}\approx0.94\,\textrm{sec},
\end{equation}
 where $\Gamma_{P_{1/2}}$ and $\Gamma_{D_{3/2}}$ are the decay rates
of the $P_{1/2}$ and the $D_{3/2}$ states respectively.

The $T_{2}$ time can potentially be reduced due to the fluctuations
of the energy shifts. We estimate the rate of relative phase fluctuations
by $\Delta(\Delta E_{1}-\Delta E_{2})\sim\frac{\Omega_{1}^{2}}{100\Omega}=0.1\,\textrm{Hz}$,
and thus the $T_{2}$ time is bounded by
\begin{equation}
T_{2}\leq\Delta(\Delta E_{1}-\Delta E_{2})^{-1}\sim10\,\textrm{sec}.
\end{equation}
As this bound is larger than the $T_{1}$ time we conclude that the
$T_{2}$ time is not reduced by fluctuations of the energy shifts.

\subsection{Polarization Errors}

Another source of noise is that of polarization imperfections. The
typical experimental error in the polarization is $\sim1\%$. This
means that $\sim1\%$ of a $\sigma^{+}$ polarized beam is in fact
$\sigma^{-}$ polarized and vice versa, causing an error within the
driving of each dark state. In order to analyze this effect we take

\begin{eqnarray}
H_{d1}^{p} & = & H_{d1}^{0}\nonumber \\
 & + & \frac{99}{100}\left\{ \Omega_{1}\cos\left[\left(\Delta+\frac{8B}{5}\right)t\right]\left(\left|p_{1}\right\rangle \left\langle d_{1}\right|+h.c.\right)+\sqrt{3}\Omega_{1}\cos\left[\Delta t\right]\left(\left|p_{1}\right\rangle \left\langle d_{3}\right|+h.c.\right)\right\} \nonumber \\
 & + & \frac{1}{100}\left\{ \sqrt{3}\Omega_{1}\cos\left[\left(\Delta\right)t\right]\left(\left|p_{1}\right\rangle \left\langle d_{1}\right|+h.c.\right)+\Omega_{1}\cos\left[\left(\Delta+\frac{8B}{5}\right)t\right]\left(\left|p_{1}\right\rangle \left\langle d_{3}\right|+h.c.\right)\right\} ,\nonumber \\
H_{d2}^{p} & = & H_{d2}^{0}\nonumber \\
 & + & \frac{99}{100}\left\{ \Omega_{1}\cos\left[\left(\Delta+\Omega+\frac{4B}{5}\right)t\right]\left(\left|p_{0}\right\rangle \left\langle d_{2}\right|+h.c.\right)+\sqrt{3}\Omega_{1}\cos\left[\left(\Delta+\Omega+\frac{12B}{5}\right)t\right]\left(\left|p_{0}\right\rangle \left\langle d_{2}\right|+h.c.\right)\right\} \nonumber \\
 & + & \frac{1}{100}\left\{ \sqrt{3}\Omega_{1}\cos\left[\left(\Delta+\Omega+\frac{12B}{5}\right)t\right]\left(\left|p_{0}\right\rangle \left\langle d_{2}\right|+h.c.\right)+\Omega_{1}\cos\left[\left(\Delta+\Omega+\frac{4B}{5}\right)t\right]\left(\left|p_{0}\right\rangle \left\langle d_{2}\right|+h.c.\right)\right\} ,\nonumber \\
\end{eqnarray}
 and proceed in the same manner as in the previous subsection. We
conclude that setting the intensity of the static magnetic field such
that the energy gap between two adjacent $D_{3/2}$ sub-levels is
$\geq10^{6}\,\textrm{Hz}$ ensures that the $T_{1}$ and $T_{2}$
times are not reduced.

\subsection{Spatial variations in the intensity of fields}

For the purpose of an interaction with a cavity mode we consider a
chain of ions which is located within the cavity. This opens the door
for another source of errors which is spatial variations in the intensity
of the fields. Spatial variations in the intensity of the dark states'
driving fields will only introduce small modifications of the energy
gaps between the dark states and the bright states, which will be
too small to cause a dephasing of the dark states. In order to analyze the effect of spatial variations
in the intensity of the static magnetic field
we consider the case where the magnitude of the magnetic field, $B$,
is changed to $B\pm\delta B$, and thus take
\begin{eqnarray}
H_{d1}^{\delta B} & = & \Delta\left|p_{1}\right\rangle \left\langle p_{1}\right|+\left(\frac{2\delta B}{5}-\frac{8B}{5}\right)\left|d_{1}\right\rangle \left\langle d_{1}\right|,\nonumber \\
 & + & \Omega_{1}\cos\left[\left(\Delta+\frac{8B}{5}\right)t\right]\left(\left|p_{1}\right\rangle \left\langle d_{1}\right|+h.c.\right)+\sqrt{3}\Omega_{1}\cos\left[\Delta t\right]\left(\left|p_{1}\right\rangle \left\langle d_{3}\right|+h.c.\right),\nonumber \\
H_{d2}^{\delta B} & = & \left(\Delta+\Omega\right)\left|p_{0}\right\rangle \left\langle p_{0}\right|+\left(\frac{6\delta B}{5}-\frac{12B}{5}\right)\left|d_{0}\right\rangle \left\langle d_{0}\right|-\left(\frac{2\delta B}{5}+\frac{4B}{5}\right)\left|d_{2}\right\rangle \left\langle d_{2}\right|\nonumber \\
 &  & +\Omega_{1}\cos\left[\left(\Delta+\Omega+\frac{4B}{5}\right)t\right]\left(\left|p_{0}\right\rangle \left\langle d_{2}\right|+h.c.\right)+\sqrt{3}\Omega_{1}\cos\left[\left(\Delta+\Omega+\frac{12B}{5}\right)t\right]\left(\left|p_{0}\right\rangle \left\langle d_{2}\right|+h.c.\right),\nonumber \\
\end{eqnarray}
 and proceed in the same manner as in the previous subsections. We
conclude that a variation in the magnitude of the magnetic field does
not introduce an energy shift to the dark state. In first order of
$\delta$, the energy shifts of the $\left|B_{i}\right\rangle $ and
$\left|C_{i}\right\rangle $ states is $\sim\delta B\ll\Omega_{1}$
. In agreement with the previous subsection, setting the intensity
of the static magnetic field such that the energy gap between two
adjacent $D_{3/2}$ sub-levels is $\sim10^{6}\,\textrm{Hz}$, and
taking care that $\frac{\delta B}{B}\sim10^{-5}$ (which we estimate
to be feasible experimentally) ensures that the $T_{1}$ and $T_{2}$
times are not reduced.

\section{Single qubit gates}

In this section we discuss the construction of single qubit gates.
As mentioned in the main text our construction allows for a \emph{direct}
rotation around one axes only. Rotations around other axes, or the
introduction of an arbitrary relative phase are possible to achieve
by other methods . In subsection A we derive the direct $\sigma_{y}$
gate. In subsection B we show how to implement a $\sigma_{x}$ gate.
In subsection C we show how to construct a $\sigma_{z}$
gate. In addition, in section \ref{QNDB} we show how an arbitrary
relative phase can be presented by utilizing a cavity-based method.

\subsection{Direct $\sigma_{y}$ gate}

We propose to implement a qubit rotation by applying a microwave field,
which is set to be on resonance with the (Zeeman) energy gap of the
$D_{3/2}$ sub-levels. Specifically, we set the microwave such that
it applies the $J_{y}$ operator (recall that $\left[H_{d},J_{y}\right]\left|D_{i}\right\rangle =0$
but $\left[H_{d},J_{x}\right]\left|D_{i}\right\rangle \neq0$ ). The
Hamiltonian of this single qubit gate is given by

\begin{eqnarray}
H_{g} & = & H_{d1}+H_{d2}\nonumber \\
 & + & i\Omega_{g}\left(\sqrt{3}\left|d_{1}\right\rangle \left\langle d_{0}\right|+2\left|d_{2}\right\rangle \left\langle d_{1}\right|+\sqrt{3}\left|d_{3}\right\rangle \left\langle d_{2}\right|\right)\cos\left[\frac{4B}{5}t\right]+h.c.\,.
\end{eqnarray}

Moving to the dark states basis in the IP and taking the RWA (with respect to the energy gap due to the zeeman splitting), we have
that
\begin{eqnarray}
H_{g}^{I} & = & \Omega_{g}\left(-\frac{3i}{2}\left|D_{2}\right\rangle \left\langle D_{1}\right|+h.c.\right)\nonumber \\
 & + & \Omega_{g}\left(-\frac{i}{4}\left|B_{2}\right\rangle \left\langle B_{1}\right|+\frac{i}{4}\left|C_{2}\right\rangle \left\langle B_{1}\right|+\frac{i}{4}\left|B_{2}\right\rangle \left\langle C_{1}\right|-\frac{i}{4}\left|C_{2}\right\rangle \left\langle C_{1}\right|\right)+h.c.\,,
\end{eqnarray}
and hence, we see that $H_{g}^{I}$ operates within the protected
subspace and corresponds to the rotation operator
\begin{equation}
\sigma_{y}^{D}=-\frac{3i\Omega_{g}}{2}\left|D_{2}\right\rangle \left\langle D_{1}\right|+h.c..
\end{equation}

\subsection{$\sigma_{x}$ gate}

In this subsection we derive an effective $\sigma_{x}$ gate. This
is achieved by introducing a second order coupling between the $\left|d_{1}\right\rangle $
and $\left|d_{2}\right\rangle $ states. We apply two control fields.
The first field corresponds to a detuned coupling of the $\left|d_{1}\right\rangle $
state to the $\left|p_{1}\right\rangle $ state with a detuning $\delta$,
and the second field corresponds to the detuned coupling of the $\left|d_{2}\right\rangle $
state to the $\left|p_{1}\right\rangle $ state with the same detuning
$\delta$. The Rabi frequency of both coupling fields is $2\Omega_{cont}$
such that $\delta\gg\Omega_{cont}.$  Denoting by $\omega_{p_{1}d_{2}}$
the rate corresponding to the energy gap between the $\left|d_{2}\right\rangle $
and the $\left|p_{1}\right\rangle $ states, and by $\omega_{1}$
and $\omega_{2}$ the frequencies of the control fields,
we begin with the Hamiltonian

\begin{eqnarray}
H_{d_{1},d_{2}} & = & \omega_{p_{1}d_{2}}\left|p_{1}\right\rangle \left\langle p_{1}\right|-\frac{4B}{5}\left|d_{1}\right\rangle \left\langle d_{1}\right|\nonumber \\
 & + & 2\Omega_{cont}\cos\left[\omega_{1}t\right]\left(\left|p_{1}\right\rangle \left\langle d_{1}\right|+h.c.\right)+2\Omega_{cont}\cos\left[\omega_{2}t\right]\left(\left|p_{1}\right\rangle \left\langle d_{2}\right|+h.c.\right).
\end{eqnarray}
Moving to the the IP with respect to $H_{d_{1},d_{2}}^{0}=\omega_{p_{1}d_{2}}\left|p_{1}\right\rangle \left\langle p_{1}\right|-\frac{4B}{5}\left|d_{1}\right\rangle \left\langle d_{1}\right|$
and making the RWA, we arrive at
\begin{equation}
H_{d_{1},d_{2}}^{I}=\Omega_{cont}\left[\left(e^{i\delta t}\left|p_{1}\right\rangle \left\langle d_{2}\right|+h.c.\right)+\left(e^{i\delta t}\left|p_{1}\right\rangle \left\langle d_{1}\right|a+h.c.\right)\right].
\end{equation}
For $\delta \gg \Omega_{cont}$ the effect of this Hamiltonian is to induce a Raman transition \cite{James}:



\begin{equation}
H_{d_{1},d_{2}}^{int}=-\frac{\Omega_{cont}^{2}}{\delta}\left(\left|d_{2}\right\rangle \left\langle d_{1}\right|+\left|d_{1}\right\rangle \left\langle d_{2}\right|\right).
\end{equation}
Moving to the dark states basis, and neglecting all terms which couple
a dark state to a non-protected state, we find that
\begin{equation}
\sigma_{x}^{D}\approx-\frac{3\Omega_{cont}^{2}}{4\delta}\left(\left|D_{2}\right\rangle \left\langle D_{1}\right|+\left|D_{1}\right\rangle \left\langle D_{2}\right|\right).
\end{equation}
Together with the above $\sigma_{y}^{D}$ gate, a $\sigma_{z}^{D}$
gate can also be implemented, and hence, any single qubit unitary
operation may be performed.

\subsection{$\sigma_{z}$ gate}

In this subsection we construct a $\sigma_{z}$ gate by employing a method for the implementation of
an adiabatic gate. The method was conceived and derived by Mikelsons {\it et al.} \cite{Gatis2013}, and
it is based on the scheme presented in \cite{Duan2001}.

Recall
that $\left|D_{1}\right\rangle =\frac{\sqrt{3}}{2}\left|d_{1}\right\rangle -\frac{1}{2}\left|d_{3}\right\rangle ,$
and define its orthogonal state
\begin{equation}
\left|D_{1}^{\perp}\right\rangle =\frac{1}{2}\left|d_{1}\right\rangle +\frac{\sqrt{3}}{2}\left|d_{3}\right\rangle .
\end{equation}
Consider the following general Hamiltonian, which is written after
performing the RWA and moving to the IP with respect to the time-independent
part,
\begin{align}
H_{ad}= & \frac{1}{2}\bigg(e^{-i\theta_{-}}\Omega_{-}\left|p_{1}\right\rangle \left\langle d_{1}\right|+e^{-i\theta_{+}}\Omega_{+}\left|p_{1}\right\rangle \left\langle d_{3}\right|+h.c.\bigg).\label{ham}
\end{align}
The adiabatic evolution can be used to construct a $\sigma_{z}$
gate. Introducing the adiabatic variables $R_{1}(t)$ and $R_{2}(t)$,
one specifies:
\begin{align}
 & \notag\theta_{+}=R_{1}\,\,,\,\,\theta_{-}=0\\
 & \notag\Omega_{-}=2\Omega_{\circ}\sin{(R_{2})}\\
 & \Omega_{+}=2\Omega_{\circ}\cos{(R_{2})}\label{c1},
\end{align}
 $\Omega_{\circ}$ fixes the adiabatic time-scale. The two
parameters $R_{1,2}(t)$ are to be varied in a closed loop, keeping
the system in the zero-eigenvalue state:
\begin{align}
\left|\Psi_{\circ}(t)\right\rangle =(0\,,\,\cos{R_{2}}\,,\,-e^{iR_{1}}\sin{R_{2}})^{T}
\end{align}
 in the basis \{$\left|p_{1}\right\rangle ,\left|d_{1}\right\rangle ,\left|d_{2}\right\rangle $\}.
The start/end point is to be fixed at ($R_{1}=0$, $R_{2}=\pi/4$)
for the $\left|D_{1}\right\rangle $- qubit.

Initial eigenstate of the system is recovered at the end of the loop
with the addition of the Berry phase, which is calculated to be:
\begin{align}
\Phi=\frac{1}{2}\iint_{{\bf \Sigma}}\sin{2R_{2}}\cdot dR_{1}dR_{2},\label{phi}
\end{align}
 where the integral is over the surface enclosed. Considering $R_{1,2}$
to be the polar co-ordinates on the unit sphere (with $2R_{2}$ in
place of $\theta$), it is seen that the Berry phase acquired will
be proportional to the solid angle swept out by the closed contour.
This procedure results in a phase which exactly corresponds to a $\sigma_{z}$
gate. Note that with the available $\sigma_{z}$ gate an arbitrary relative phase can be introduced.


\section{Interaction with a cavity mode}

In this section we derive the Hamiltonian of the effective coupling
between a cavity mode and a protected qubit. This is achieved by setting
the cavity mode such that its frequency and polarization correspond
to the detuned coupling of the $\left|d_{1}\right\rangle $ state
to the $\left|p_{1}\right\rangle $ state with a detuning $\delta$
. In addition, we apply an external control field which corresponds
to the detuned coupling of the $\left|d_{2}\right\rangle $ state
to the $\left|p_{1}\right\rangle $ state with the same detuning $\delta$
and with a Rabi frequency $2\Omega_{cont}$ such that $\delta\gg\Omega_{cont}\gg g$,
where $g$ is the rate describing the coupling between a single photon
in the cavity mode to a single ion (see Fig. 3 in main text). Denoting
by $\omega_{p_{1}d_{2}}$ the rate corresponding to the energy gap
between the $\left|d_{2}\right\rangle $ and the $\left|p_{1}\right\rangle $
states and by $\omega_{cont}=\omega_{p_{1}d_{2}}-\delta$ the frequency
of the control field, we begin with the Hamiltonian

\begin{eqnarray}
H_{i,c} & = & \omega_{c}a^{\dagger}a+\omega_{p_{1}d_{2}}\left|p_{1}\right\rangle \left\langle p_{1}\right|-\frac{4B}{5}\left|d_{1}\right\rangle \left\langle d_{1}\right|\nonumber \\
 & + & 2\Omega_{cont}\cos\left[\omega_{cont}t\right]\left(\left|p_{1}\right\rangle \left\langle d_{2}\right|+h.c.\right)+g\left(\left|p_{1}\right\rangle \left\langle d_{1}\right|a+h.c.\right).
\end{eqnarray}
Moving to the the IP with respect to $H_{i,c}^{0}=\omega_{c}a^{\dagger}a+\omega_{p_{1}d_{2}}\left|p_{1}\right\rangle \left\langle p_{1}\right|-\frac{4B}{5}\left|d_{1}\right\rangle \left\langle d_{1}\right|$
and making the RWA, we arrive at

\begin{equation}
H_{i,c}^{I}=\Omega_{cont}\left(e^{i\delta t}\left|p_{1}\right\rangle \left\langle d_{2}\right|+h.c.\right)+g\left(e^{i\delta t}\left|p_{1}\right\rangle \left\langle d_{1}\right|a+h.c.\right).
\end{equation}
For $\delta \gg \Omega_{cont} \gg g$ the effect of this Hamiltonian is to induce a Raman transition \cite{James}:

\begin{equation}
H_{i,c}^{int}=-\frac{g\Omega_{cont}}{\delta}\left(\left|d_{2}\right\rangle \left\langle d_{1}\right|a+\left|d_{1}\right\rangle \left\langle d_{2}\right|a^{\dagger}\right).
\end{equation}
Moving to the dark states basis, and neglecting all terms which couple
a dark state to a non-protected state, we find that

\begin{equation}
H_{i,c}^{int}\approx-\frac{3g\Omega_{cont}}{4\delta}\left(\left|D_{2}\right\rangle \left\langle D_{1}\right|a+\left|D_{1}\right\rangle \left\langle D_{2}\right|a^{\dagger}\right).
\end{equation}

\section{Quantum non-demolition measurement}

In this section we show how to perform a quantum non-demolition measurement
(QND) of the photon-number in the cavity. We derive the QND Hamiltonian
in subsection A. Then, in subsection B we note that the same Hamiltonian
can be used in order to present an arbitrary relative phase to a protected
qubit.

\subsection{Quantum non-demolition measurement}

Removing the control field in the setup described in the previous
section, we are left only with the coupling of the ions to a cavity
mode. The Hamiltonian of a single ion and a cavity mode is then given
by

\begin{eqnarray}
H_{i,c} & = & \omega_{c}a^{\dagger}a+\omega_{p_{1}d_{2}}\left|p_{1}\right\rangle \left\langle p_{1}\right|-\frac{4B}{5}\left|d_{1}\right\rangle \left\langle d_{1}\right|+g\left(\left|p_{1}\right\rangle \left\langle d_{1}\right|a+h.c.\right).
\end{eqnarray}
Moving to the the IP with respect to $H_{i,c}^{0}=\omega_{c}a^{\dagger}a+\omega_{p_{1}d_{2}}\left|p_{1}\right\rangle \left\langle p_{1}\right|-\frac{4B}{5}\left|d_{1}\right\rangle \left\langle d_{1}\right|$
and making the RWA, we arrive at

\begin{equation}
H_{i,c}^{I}=g\left(e^{i\delta t}\left|p_{1}\right\rangle \left\langle d_{1}\right|a+h.c.\right).
\end{equation}
For $\delta \gg g$ the effect of this Hamiltonian is to induce a Stark shift \cite{James}:
\begin{equation}
H_{i,c}^{int}=\frac{g^{2}}{\delta}\left(\left|p_{1}\right\rangle \left\langle p_{1}\right|-\left|d_{1}\right\rangle \left\langle d_{1}\right|\right)a^{\dagger}a.
\end{equation}
Moving to the dark states basis, and neglecting all terms which are
outside of the protected subspace or couple states within within the
protected subspace to states outside the protected subspace, we have
that

\begin{equation}
H_{i,c}^{int}\approx-\frac{3g^{2}}{4\delta}\left|D_{1}\right\rangle \left\langle D_{1}\right|a^{\dagger}a.
\end{equation}

As $H_{i,c}^{int}$ takes $\frac{1}{\sqrt{2}}\left(\left|D_{2}\right\rangle +\left|D_{1}\right\rangle \right)$
to $\frac{1}{\sqrt{2}}\left(\left|D_{2}\right\rangle +e^{i\frac{3g^{2}ta^{\dagger}a}{4\delta}}\left|D_{1}\right\rangle \right)$,
a non-demolition measurement of the photon-number in the cavity can
be performed by a Ramsey spectroscopy experiment on the dark states.

\subsection{Arbitrary cavity-based phase of a protected qubit}\label{QNDB}

The above observation can be used in order to introduce an arbitrary
phase to a protected qubit. After setting the amplitudes of the qubit,
$\left|\psi\right\rangle =a\left|D_{1}\right\rangle +b\left|D_{2}\right\rangle $,
an arbitrary relative phase cab be added by placing the ion in a cavity
with a known number of photon. After the interaction the state of
the qubit is given by
\begin{equation}
\left|\psi\right\rangle =a\left|D_{1}\right\rangle +be^{-i\varphi}\left|D_{2}\right\rangle ,
\end{equation}
 where $\varphi=\frac{3g^{2}ta^{\dagger}a}{4\delta}$. The phase can
therefore be controlled by the number of photons, the duration time
of the interaction, and by the frequency of the mode.


\end{widetext}


\begin{references}

\bibitem{Benhelm2008} J. Benhelm \emph{et al.},
\href{http://pra.aps.org/abstract/PRA/v77/i6/e062306}{Phys. Rev. A \textbf{77}, 062306 (2008)}.


\bibitem{Lucas2007} D. M. Lucas \emph{et al.}, \href{http://arxiv.org/abs/0710.4421}
{e-print arXiv:0710.4421}.

\bibitem{Olmschenk2007} S. Olmschenk \emph{et al.}, \href{http://pra.aps.org/abstract/PRA/v76/i5/e052314}{Phys. Rev. A \textbf{76},
052314 (2007)}.

\bibitem{Langer2005}
C. Langer \emph{et al.},
\href{http://prl.aps.org/abstract/PRL/v95/i6/e060502}{Phys. Rev. Lett. \textbf{95}, 060502 (2005)}.

\bibitem{rare1}  E. Fraval,
\emph{et al.}, \href{http://prl.aps.org/abstract/PRL/v92/i7/e077601}{Phys. Rev. Lett. {\bf 92}, 077601 (2004)}.

\bibitem{Lidar1998}
D. A. Lidar \emph{et al.},
\href{http://prl.aps.org/abstract/PRL/v81/i12/p2594_1}{Phys. Rev. Lett. {\bf 81}, 2594 (1998)}.

\bibitem{Kielpinski2001}
D. Kielpinski \emph{et al.},
\href{http://www.sciencemag.org/content/291/5506/1013.short}{Science \textbf{291},  1013 (2001)}.

\bibitem{Haeffner2005}
H. H\"affner \emph{et al.},
\href{http://link.springer.com/article/10.1007/s00340-005-1917-z}{Applied Physics B \textbf{81},  151 (2005)}.

\bibitem{Viola1998}
L. Viola and S. Lloyd, \href{http://pra.aps.org/abstract/PRA/v58/i4/p2733_1}{Phys. Rev. A \textbf{58}, 2733 (1998)}.

\bibitem{Biercuk2009} Michael J. Biercuk \emph{et al.},
\href{http://www.nature.com/nature/journal/v458/n7241/full/nature07951.html}{Nature  \textbf{458}, 996 (2009)}.

\bibitem{Naydenov2011} B. Naydenov \emph{et al.},
\href{http://prb.aps.org/abstract/PRB/v83/i8/e081201}{Phys. Rev. B \textbf{83}, 081201(R) (2011)}.

\bibitem{Tan2013}
T. R. Tan {\it et al.}, \href{http://arxiv.org/abs/1301.3786}{arXiv:1310.3786}.

\bibitem{Goldman}
M. Goldman, Spin Temperature and Nuclear Magnetic Resonance in Solids, Clarendon Press, Oxford, 1970.

\bibitem{CaiCon}
J-M Cai {\it et al.},
\href{http://iopscience.iop.org/1367-2630/14/11/113023/article}{New J. Phys. \textbf{14} 113023 (2012)}.

\bibitem{NVGate}
A. Bermudez {\it et al.},
\href{http://prl.aps.org/abstract/PRL/v107/i15/e150503}{Phys. Rev. Lett. \textbf{107}, 150503 (2011)}.

\bibitem{CaiCavity}
Jianming Cai {\it et al.},
\href{http://iopscience.iop.org/1367-2630/14/9/093030?fromSearchPage=true}{New J. Phys. \textbf{14} 093030}.


\bibitem{Christof}
N. Timoney {\it et al.}, \href{http://www.nature.com/nature/journal/v476/n7359/full/nature10319.html}{Nature \textbf{476}, 185 (2011)}.

\bibitem{norm_condition}
The actual condition is that the lowest eigenvalue of $H_d$ of an eigenstate which is outside the kernel is much larger than the characteristic frequency of the noise. In this case the new $T_1$ time would be proportional to the power spectrum at a frequency which equals the smaller eigenvalue.

\bibitem{knill}
E. Knill {\it et al.},
\href{http://www.fas.org/sgp/othergov/doe/lanl/pubs/00783364.pdf}{Los Alamos Sci. \textbf{27}, 188 (2002)}.

\bibitem{Webster2013}
S. C. Webster {\it et al.},
\href{http://prl.aps.org/abstract/PRL/v111/i14/e140501}{Phys. Rev. Lett. \textbf{111}, 140501 (2013)}.

\bibitem{Gatis2013}
Mikelsons {\it et al.}, to be published.

\bibitem{supp}
See Supplementary Material.

\bibitem{laser}
However, it is possible to keep the intensity of the laser beams stable at a level considerably below one percent, including at frequencies of 1 Hz or below, by the use of commercial power stabilisers (e.g., \href{http://www.thorlabs.com/}{ThorLabs Liquid crystal Noise Eaters}).

\bibitem{T2_limit}
The actual $T_2$ due to a Markovian noise is $\frac{1}{g^2  \tau_c}$, where $g$ is the rate of the noise and $\tau_c$ is the correlation time of the noise. A lower bound on $T_2$ is given by $1/g$, which is achieved for a long $\tau_c$ ($\tau_c=T_2$).

\bibitem{Herskind2009}
P. F. Herskind {\it et al.},
\href{http://www.nature.com/nphys/journal/v5/n7/full/nphys1302.html}{Nat. Phys. \textbf{5}, 494 (2009)}.


\bibitem{Albert2011}
M. Albert {\it et al.},
\href{http://www.nature.com/nphoton/journal/v5/n10/abs/nphoton.2011.214.html}{Nat. Phot. \textbf{5}, 633 (2011)}.







\bibitem{Keller2007}
M. Keller {\it et al.},
\href{http://www.tandfonline.com/doi/full/10.1080/09500340600792911#.Uclv0js3DTq}{Jour. Mod. Opt. \textbf{54}, 1607 (2007)}.



\bibitem{Dubin2010}
F. Dubin {\it et al.},
\href{http://www.nature.com/nphys/journal/v6/n5/full/nphys1627.html}{Nat. Phys. \textbf{6}, 350 (2010)}.

\bibitem{Stute2012}
A. Stute {\it et al.},
\href{http://www.nature.com/nature/journal/v485/n7399/full/nature11120.html}{Nature \textbf{485}, 482 (2012)}.

\bibitem{Stute2013}
A. Stute {\it et al.},
\href{http://www.nature.com/nphoton/journal/v7/n3/abs/nphoton.2012.358.html}{Nat. Phot. \textbf{7}, 219 (2013)}.


%
%


\bibitem{galaxy}
L. Hornek¾r{\it et al.},
\href{http://prl.aps.org/abstract/PRL/v86/i10/p1994_1}{Phys. Rev. Lett. \textbf{86}, 1994 (2001)}.

\bibitem{Clausen2013}
C. Clausen {\it et al.},
\href{http://iopscience.iop.org/1367-2630/15/2/025021}{New. J. Phys. \textbf{15}, 025021 (2013)}.










\end{references}

\begin{references}

\bibitem{laser}
However, it is possible to keep the intensity of the laser beams stable at a level considerably below one percent, including at frequencies of 1 Hz or below, by the use of commercial power stabilizers (e.g., \href{http://www.thorlabs.com/}{ThorLabs Liquid crystal Noise Eaters}).

\bibitem{James}
D. F. James and J. Jerke,
Canadian Journal of Physics, \textbf{85}, 625 (2007).

\bibitem{Gatis2013}
Mikelsons {\it et al.}, to be published.

\bibitem{Duan2001}
L.-M. Duan {\it et al.}, Science \textbf{292}, 1695 (2001).



\end{references}
\end{document}